# "Identification of single events in the HPGe detector: Comparison of various methods based on the analysis of simulated pulse shapes"

Bakalyarov A.M., Balysh A.Ya., Belyaev S.T., Lebedev V.I., Zhukov S.V.

## SUMMARY

Various methods of identification of single events in the HPGe detector are compared on the basis of a program especially designed to simulate pulse shape in a semi-conductor germanium detector. Capabilities of three following methods are shown: (1) an application of the library of single pulse shapes, (2) the single-parameter method, and (3) a separation on the basis of artificial neural networks. The analysis was done in the context of an application of all above-mentioned techniques in the energy interval around the expected neutrino less double beta decay of Ge-76 in the Heidelberg-Moscow experiment.

# 1. Introduction

One of the primary goals in the study of the double beta decay of Ge-76 in the Heidelberg-Moscow experiment is a background reduction in the energy interval of the neutrinoless (0νββ) mode [1]. One of possible means for this reduction employs a division of background events into two classes (single and multiple ones) on the basis of an analysis of a pulse shape in the HPGe detector with subsequent suppression of multiple events which are almost completely of a background nature.

At the moment, several methods of a solution of this problem are known: a compilation of the library of single event shapes [2], the single-parameter method for single events identification [3], and an application of artificial neural networks [4]. Every above-mentioned technique is far from an ideal. There exists a rather high probability to identify a single event as a multiple one (an **α**-probability), and vice versa, to identify a multiple event as a single one (a **β**-probability). This fact results either in an omission of an effect or in a non-effective background suppression. These probabilities are interconnected – if one of errors would be decreased then the other would be increased. Unfortunately, under practical application of all identification techniques, an exact determination of these probabilities is connected with inherent difficulties. It is rather problematic to get an experimental set of pulses including only single events or only multiple events and having α- and β-probabilities to be equal to values typical for an experiment in the energy interval around the expected 0νββ signal. It seems to be incorrect if one would use energy peaks in testing: while these peaks consist mostly of events belonging to a certain class, they have a significant part of events belonging to another class as well. Joint testing of two different peaks, for example, the peak from the gamma line of 2614 keV and the "double escape" peak (1592 keV) for the same primary line (Th-228), would be correct if only an **α**-value (as well as a **β**-value) is the same for both peaks. Moreover, results of this testing could be applied to the energy interval of the 0νββ-mode if only values of **α**- and **β**-errors inside peaks would be equal to those over the energy interval being of interest in the real double-beta-decay experiment. A situation is arisen when test of various methods for single events identification becomes a very important problem, indeed. Incorrect determination of errors could result in a significant uncontrolled loss of real events.

We do not see any other possibility now except an application of computer simulation for a solution of this problem. A simulation provides a chance both to make qualitative estimation of errors and to study their dependence on factors being of interest to us. The main goal of the presented paper is an investigation of all three above-mentioned techniques and a comparison of corresponding results when a simulation of pulse shapes from 0νββ electrons and from the gamma source of Th-228 was given as an example.

## 2. Simulation of a pulse shape

In order to get a pulse shape in the HPGe detector one needs the following:
- To simulate an interaction of a primary radiation (gamma quantum or electron) with the detector and to locate a point of an initial origin of electron-hole pairs in the semiconductor detector.
- To calculate an amplitude of an electric field at each point inside the detector as a function of an applied external potential taking into account a given amount and a spatial distribution of impurities concentrations.
- To determine a shape of a current pulse for a calculated electric field distribution and for initial points of holes and electrons origination.
- To get a shape of an output voltage pulse taking into account a positioning of the detector into a real electric circuit with known parameters.

Similar models were studied elsewhere [e.g., 5-7]. We have designed our own original programs. For simulation, one of detectors was chosen (Detector 5) which is located now inside the operational experimental setup for an investigation of the double beta decay in the Heidelberg-Moscow experiment. The detector was of 78.8 mm in diameter and 105.7 mm in length.

An interaction of a primary radiation with the detector as well as a process of propagation and detection of particles was studied with the help of well-known program Geant3 [8] with a real geometry of the detector and its location inside the experimental setup taken into account. Coordinates of origination points for electron-hole pairs as well as a value of the energy absorbed in the form of the same electrons and holes were stored in a memory in a process of simulation for each primary gamma quantum or electron. These data were written into the special "event" file, which was subsequently used for a simulation of a sequence of pulse shapes in the detector. Two kinds of event files were obtained: for simulation of 0νββ electrons and of gamma quanta from the Th-228 source. Electrons from the 0νββ mode were originated uniformly inside the detector. Their initial energetic and angular distributions were simulated on the basis of known algorithms [9]. The Th-228 source was positioned into a center of the experimental setup for a simulation; it used to be at the same place in the process of calibration [10].

Electric field inside the detector was computed by solution of the Poisson equation

$$\Delta\Phi = -\rho/\varepsilon \qquad (1)$$

where $\Delta\Phi$ is the Laplacian of a potential, $\rho$ is a distribution function for free charges, and $\varepsilon$ is a dielectric constant of germanium.

A cylindrical symmetry of the detector reduces the differential equation down to two-dimensional problem, which was solved by using a known technique [10]. It should be noted that a spatial interval of a computational lattice was equal to 0.1 mm. As a result, we got a file providing information on electric potential distribution inside the detector with the spatial resolution of 0.1 mm. The electric field distribution inside the detector is presented in Fig.1. It was assumed that free carriers are uniformly distributed inside the detector and their concentration is of $5 \cdot 10^9 cm^{-3}$.

A simulation of a current signal shape in the HPGe detector is executed by a program, which employs a given distribution of initial points of electron-hole pairs origination (taken for every pulse from the event file) and a given distribution of the electric field (taken from the calculated potential-distribution file). A current signal is formed as a result of charge carriers drift to electrodes. The current amplitude is a function of the electric field **E** and the drift velocity **V** along the electric field line inside the detector. A pulse shape was simulated by a successive computation of the amplitude and the direction of the electric field when a motion of electrons and holes to electrodes was considered within every temporal interval of 1 ns. Values for electrons and holes drift velocities as well as their dependence on the electric field and a carrier mobility were taken from [10].

A simulation takes into account the fact that the detector is a part of an electric circuit. Corresponding equivalent circuit consists of a source of current (current pulse from the detector), detector's resistance and capacity, load's resistance and capacity, and transitional equivalent capacity and resistance (see Fig.2). An application of an equivalent circuit's concept allows to get pulse shape which is nearly similar to an experimental one. The output signal is measured on the load resistance $R_L$. A computation of the output signal was carried out by a numerical solution of the system of differential equations with the help of the Runge-Kutta method. A temporal interval in the process of numerical integration was of $10^{-12}$ s. It should be noted that a program that simulates a current signal in the detector as well as integration by the Runge-Kutta method are a single whole: an integration of a pulse shape on the load resistance is executed over time. Fig.3 shows some typical pulses which were computed according to simulation of 0νββ electrons.

## 3. Identification of single events with the help of the library of single event shapes.

We have compiled the theoretical library of single event shapes in order to understand capabilities and limitations of the proposed method [2]. This library is a set of a definite amount of simulated pulse shapes. It was assumed in the process of computing that all interaction energy is deposited at a single point inside the detector. Point's distribution was chosen as uniform one over the detector's volume. In order to compile this library we have simulated 10 000 pulses of such a kind.

We have simulated two groups of pulses: pulses from the 0νββ mode of Ge-76 decay with the total energy of 2038 keV (two electrons originated at a single point), and pulses from gamma quanta of Th-228 with the total energy of 2000-2100 keV absorbed inside the detector. It was done in order to get a qualitative estimate of a possible background suppression factor in the energy interval of the 0νββ mode when using the method under consideration. These two groups of pulses (1000 pulses in each group) were formed on the base of event files according to the procedure described in the previous Section. A shape of every pulse from the group was compared in turn with shapes of all pulses from the library of single pulses: the most likely shape of a library pulse was looked for. In other words, we have looked for the minimum of the following function:

$$S(min) = MIN\{\Sigma[A(i) – A(i,j,lib)] \; i=1…n; \; j=1…10000\} \qquad (2)$$

where n – amount of points for each pulse, and j – an index of a library pulse.

Probability functions S(min) were plotted for 0νββ electrons and for Th-228 background in the energy interval of 2000-2100 keV. Corresponding integral distributions are demonstrated in Fig.4. If one would select events with S(min) values below some predetermined threshold and reject events with S(min) values above this threshold then a detection of gamma quanta from Th-228 could be suppressed rather effectively. These quanta imitate background events in the energy interval typical for the 0νββ mode; if they would be rejected then an efficiency of 0νββ electrons detection would not be decreased in a significant manner. It is evident, that if a non-efficiency of 0νββ electrons detection is of 20%, then the background level would be decreased approximately by 2.5 times. It is a rather high value of the background suppression factor if one would take into account also that the Th-228 background in the energy interval of 2000-2100 keV consists of single events (30%) and of multiple events (70%). So, an amount of multiple events per se at this threshold value is reduced by a factor of 4.5. It is clearly seen in Fig.5 where integral distributions S(min) are presented separately for multiple and single events.

Any improvement of these results is impossible from practical point of view. We'll try now to explain this situation. The library consisting of 10 000 points which are uniformly distributed over the volume of the detector would correspond to a mean distance between points less than 1 mm. If the total energy of electrons originated due to their interaction with the detector would be deposited at a point then the quality of identification might be made better by increasing a size of the theoretical library, that is by lowering a mean distance between points. In fact, an electron event is spread over some area. We have calculated a spatial distribution of points where electrons deposit their energy. It was done by a processing of the event file for a case when this file corresponded to a simulation of 0νββ electrons. Fig.6 shows a probability distribution of maximum distance between points where electron deposited any part of its energy. It is evident that more than 60% of events have a spatial spread of more than 0.5 mm, 35% - more than 1mm, and 12% - more than 2 mm (let us remind that the electron with the energy of 2 MeV has a range of about 1.5 mm in germanium). In other words, a manifold of pulse shapes from electrons does not reduce to a pulse shape corresponded to a deposit of the total energy at a single point of the detector. The power of the set that consists of pulse shapes from real electron events is higher than the power of the set of single point events. Therefore, there exists a pure physical reason for explanation of the fact why further improvement of characteristics of electron events identification is impossible. This reason is connected with a considerable spatial spread ("spot size") of an area where electrons deposit their energy inside the detector.

### 4. The single-parameter method of an identification of single events.

Let us consider now the method of an identification of single events [3]. This method takes into account the single parameter designated as "broad" and expressed as a ratio of a maximum pulse value to a minimum of its second derivative:

$$\text{"Broad"} = \max P / \min [d^2 P / dt^2] \qquad (3)$$

All events with this parameter's value below some definite threshold are considered as single ones; if the threshold is overcome they are considered as multiple ones. We have processed simulated pulses from two groups (0νββ electrons and gamma quanta from Th-228 in the energy interval of 2000-2100 keV) with the help of the single-parameter algorithm. For each group, 5000 events were sampled. Fig.7 demonstrates integral distributions for the probability function when both groups have a given value of the "broad" parameter. It is evident that if a non-efficiency of detection for 0νββ electrons is of 20% then the factor of background suppression would be of 1.5. Fig.8 shows similar distributions separately for single and multiple events from the second group (2000-2100 keV from Th-228). The multiple-events suppression factor in the background region would be only of 1.5 at the same threshold value.

### 5. An application of artificial neural networks.

Main ideas of single-events identification with the help of artificial neural networks are stated elsewhere [4]. One needs a set of "pure" classes of pulses to educate a neural network, that is, one needs to have a separate set of pulses from definitely single events and another separate set of pulses consisted only of multiple events.

In the process of a simulation there are no problems to get any set of pulses. However, one has to take unto account that the method should have a practical application. In other words, one has to be able to get in an experimental way single and multiple pulse classes separately. Strictly speaking, it is impossible to get in the experimental way *[)] "pure" classes of sets. Only in the course of special and lengthy experiments (see, for example, [2]) one could get a good representative sampling. It would be much more difficult to do this job if the detector is a part of the operating experimental device. One of chances to provide education is connected with a use of a specific pulse group where representatives of one class do prevail. We have chosen for an education pulses in the two-peaks energy region for the Th-228 source: the first peak corresponds to the total absorption of the 2614 keV-line, the second peak of 1592 keV corresponds to the double escape process connected with the 2614 keV-line. If an energy resolution is taken unto account then the first peak has 94% of multiple events and the second peak has 78% of single events. These values were obtained on the basis of Monte-Carlo simulation.

Calculations were done with the help of the Geant 321 program. We have chosen for an education a neural network, namely, four-layer perceptron consisting of 40 input neurons, one output neuron and two hidden neuron layers (15 and 8 neurons). We have formed for an education a special file consisting of 10 000 events (a half of all events was for the 2614 keV-peak and the other half – for the 1592 keV-peak. Each event was written into the file with 41 digits. Forty digits were used for a normalized pulse shape (19 points to the left from the maximum, the maximum, and 20 points to the right); the last digit was "1" (for the 1592 keV-peak) or "0" (for the 2614 keV-peak). A temporal interval for any point of a pulse shape was equal to 10 ns. The education was done with the help of the JETNET 3.0 package [14]. In the process of education, the file with 10 000 events was presented up to 5000 times.

A testing was performed with electrons of the $0\nu\beta\beta$ mode of Ge-76 decay (a total energy of 2038 keV) and with gamma quanta of the Th-228 source within the energy range of 2000-2100 keV. For these groups files were formed similar to those for education. Testing has shown that a suppression factor for the Th-228 spectrum is about two when a non-efficiency of 0νββ electrons detection is of 20%. An amount of multiple events in the spectrum is decreased in this case more than three times. Fig.9 shows a dynamics of variation of tested parameters (a share of single events for 0νββ electrons and for the source spectrum) depending on an education period for the neural network at a given testing threshold (0.5). A differential probability density of the perceptron's output signal for both 0νββ electrons and the Th-228 quanta in the energy region of 2000-2100 keV are plotted in Fig.10. Fig.11 demonstrates similar distributions for single and multiple events separately.

---

*[)] Not going into details, we indicate two aspects, which show that one, can't get a class of a 100%-purity. As Fig.6 demonstrates, a distance between two points where an electron deposits its energy may exceed appreciably the maximum electron's range in germanium. In particular, a long tail of this distribution is connected with bremsstrahlung gamma quanta; therefore, a real electron event ($0\nu\beta\beta$) is transformed into a category of multiple events. On the other hand, the gamma quantum may deposit its energy at two points of the detector, which are separated appreciably from each other but lie on the same equipotential surface. In this case, a real multiple event would be undistinguished from a single one with energy equal to the total energy losses by the gamma-quantum.

A testing over the whole spectral region (400-2300 keV) from the Th-228 source is of independent interest. In the process of testing we did not consider areas under peaks. Fig.12 presents an energy dependence of a ratio of total perceptron output (with a 0.5 threshold) to a total amount of events at the input of the neural network. Two curves have to do with single and multiple events, respectively. It is seen that if no more than 20-25% of single events are omitted then multiple events are suppressed by 3-5 times.

It is of interest to compare two sets of libraries, namely, the library of single pulse shapes
(Employed in Section 3), and the library employed for the method of artificial neural networks. As mentioned before, a temporal interval during a computing of a pulse shape was equal to 1 ns in the first case and to 10 ns in the second one. The second library was used in order to get the integral probability function S(min) similar to that in Section 3. Corresponding results are plotted in Figs.13-14.

In spite of the fact that a scale along the X axes has changed practical results were not changed at all. If a non-efficiency of 0νββ electrons detection is about 20% then the background value is decreased by 2.5 times and multiple events are suppressed by 4.5 times.

## 6. Conclusion

We have compared on the simulation basis three various methods of single (and multiple) events identification with the help of pulse shape analysis. The method with the library of single pulses results in a satisfactory background suppression. Simple estimations show that this method can make a half-life limit for the 0νββ mode of Ge-76 decay more accurate (by 25-30%) if compared with a half-life value obtained
on the basis of background estimation without taking pulse-shape analysis into account. **[)] This is probably an ultimate improvement, which can be done on the basis of a pulse-shape analysis. More exact evaluation might be obtained with the help of more precise simulation of all background sources which are important in the energy region relevant to the 0νββ mode of Ge-76 decay. It should be noted that a compilation of the library of pulse shapes for single events is a very complicated problem per se. We have assumed in our analysis that this complete library does exists already. In fact, the process of the library formation (both experimental or theoretical ones) might be accompanied by additional errors, which were not considered in our analysis.

---

**[)] Here and further, when estimating the period of half-life, it was assumed, that the structure of the background (quantity of single and multiple events in experimental spectrum) was just like the structure in Compton part of the spectrum from Th-228 source.

The single-parameter method of single-event identification does not provide more accurate definition of a half-life limit for the 0νββ mode. It is connected with a low value of the multiple-events-suppression factor. But the basic reason is that no analysis of the whole pulse shape was implied.

The identification method based upon an education of a neural network provides quite satisfactory results. It allows to increase a limit on a half-life of Ge-76 decay in the 0νββ mode by 10-15% without any additional analyses or experiments. This method provides a satisfactory suppression of multiple events as well.

Results of an application of neural networks are not worse than results of analysis with an employment of library pulses. However, a need in classes of sets (which are as "pure" as possible) for network education reverts us to a problem of special lengthy experiments (which were described in details in [3]). In the case of the theoretical library compilation one needs to know detector parameters as precisely as possibly including impurities concentrations and their distribution over the detector volume.

## Acknowledgements

We would like to thank Dr. B.Majorovits for useful discussions and help in the experiment for development of pulse-shape method.

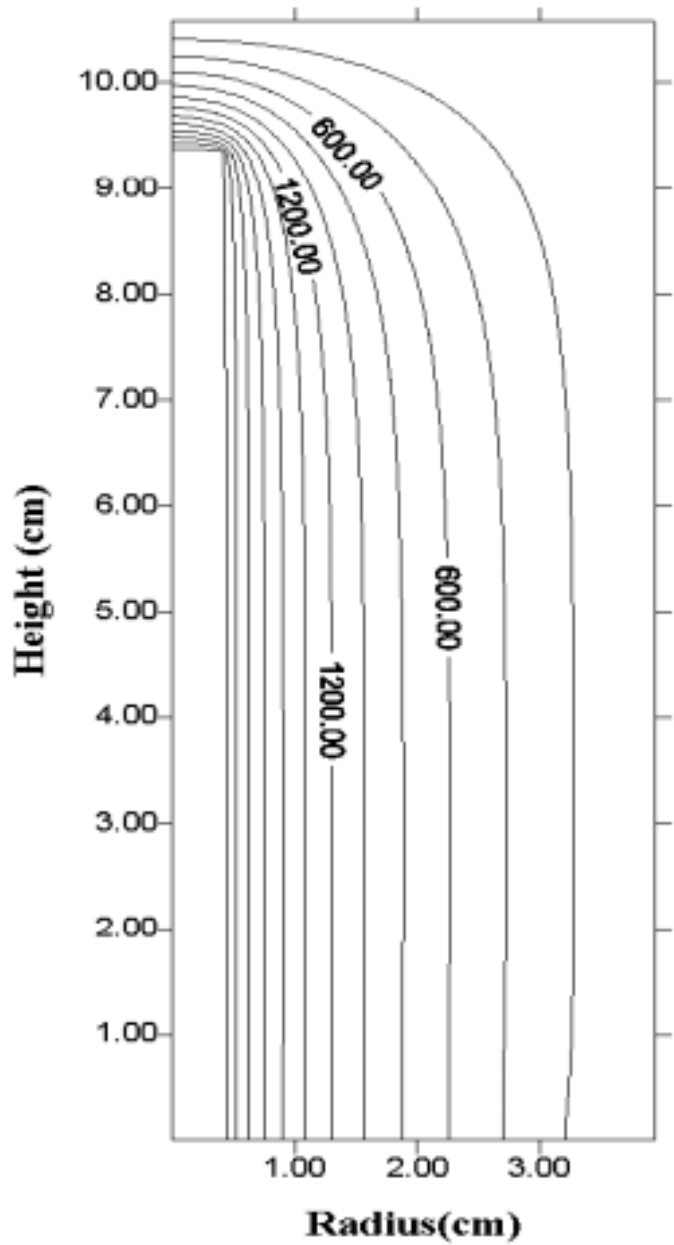

Fig.1. Electrical potential distribution in detector

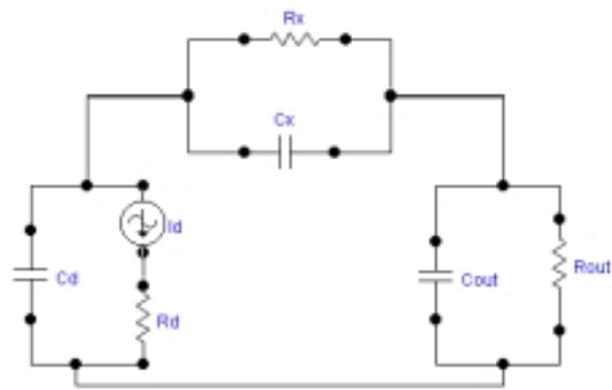

Fig.2. Electrical circuit for pulse shape calculation

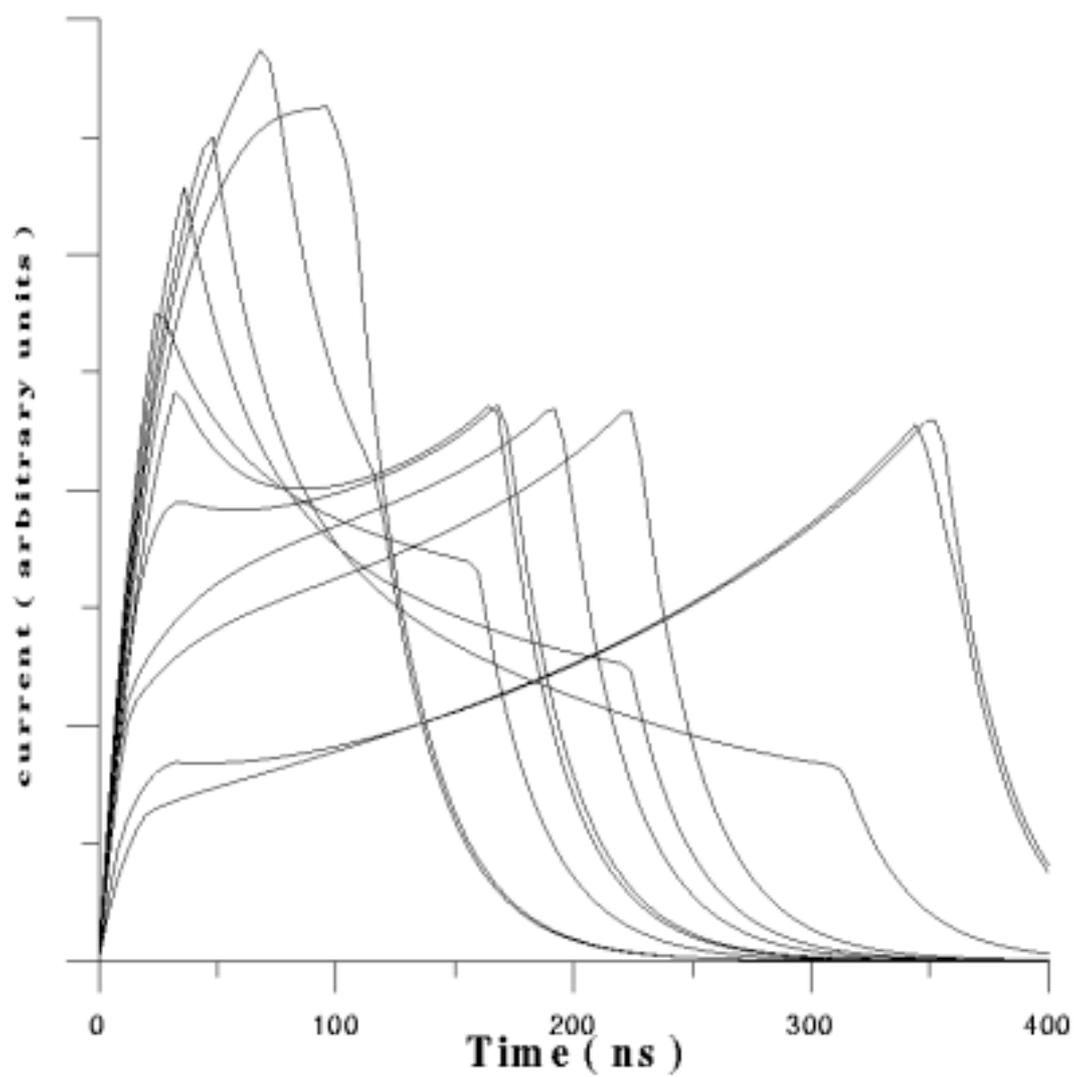
Fig.3.Simulated pulse shapes for double beta decay electrons

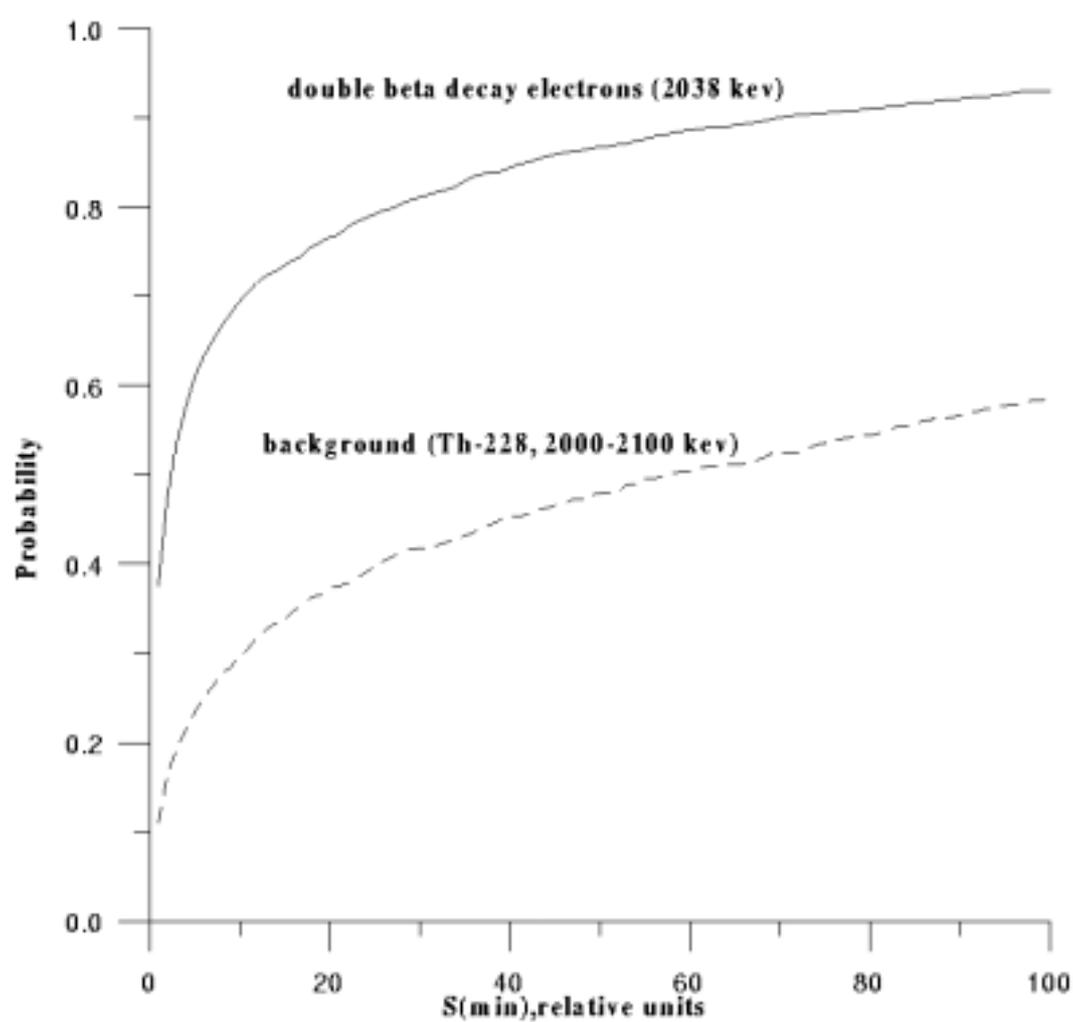

Fig.4. Integral density of S(min)-probability

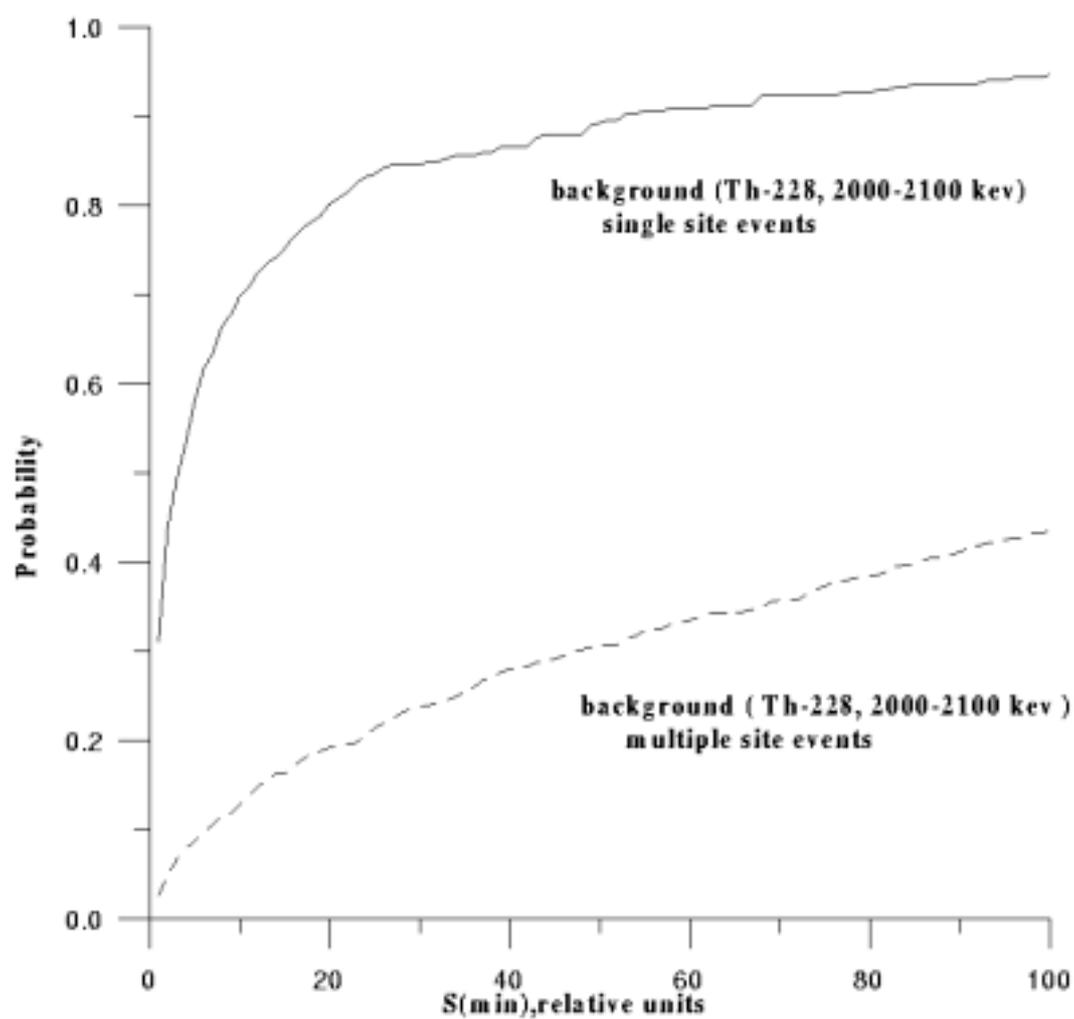

Fig.5. Integral density of S(min)-probability

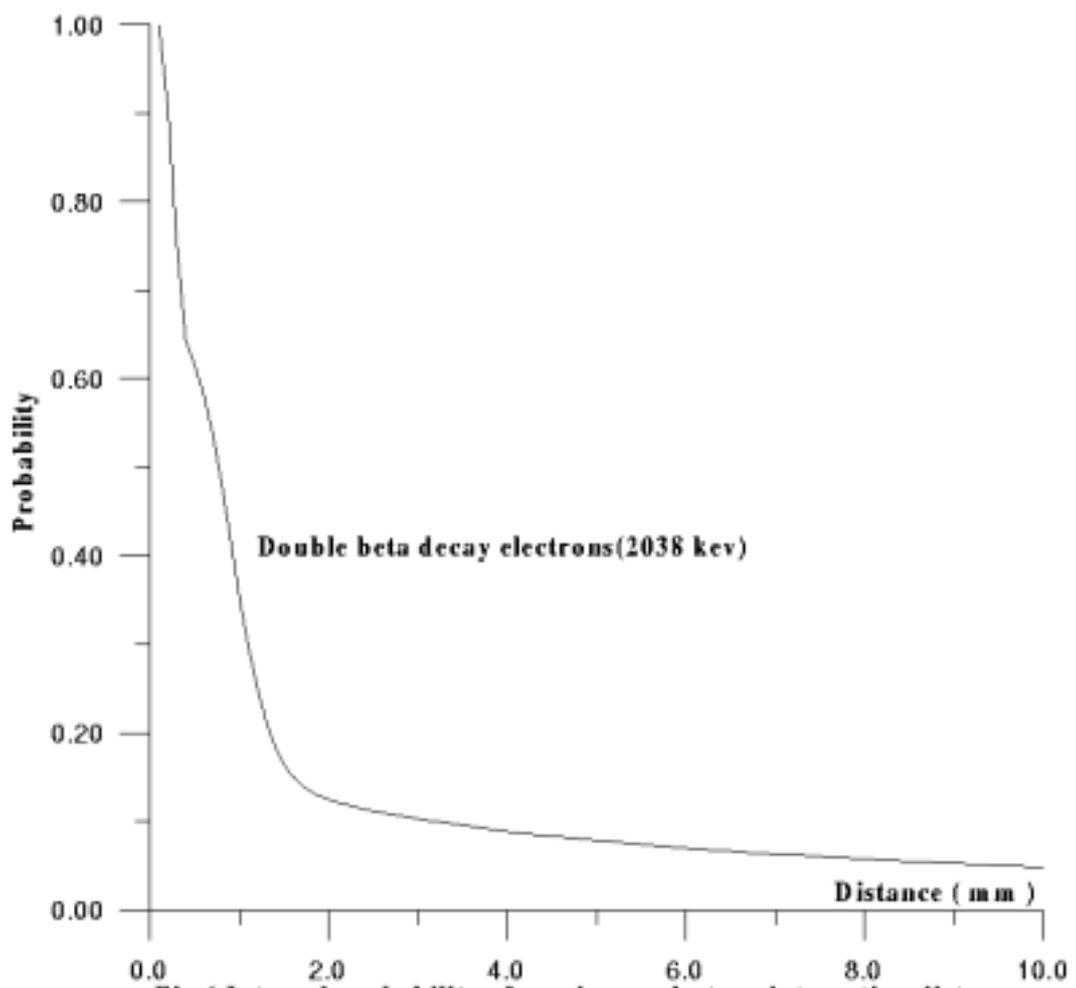
Fig.6 Integral probability of maximum electron interaction distance

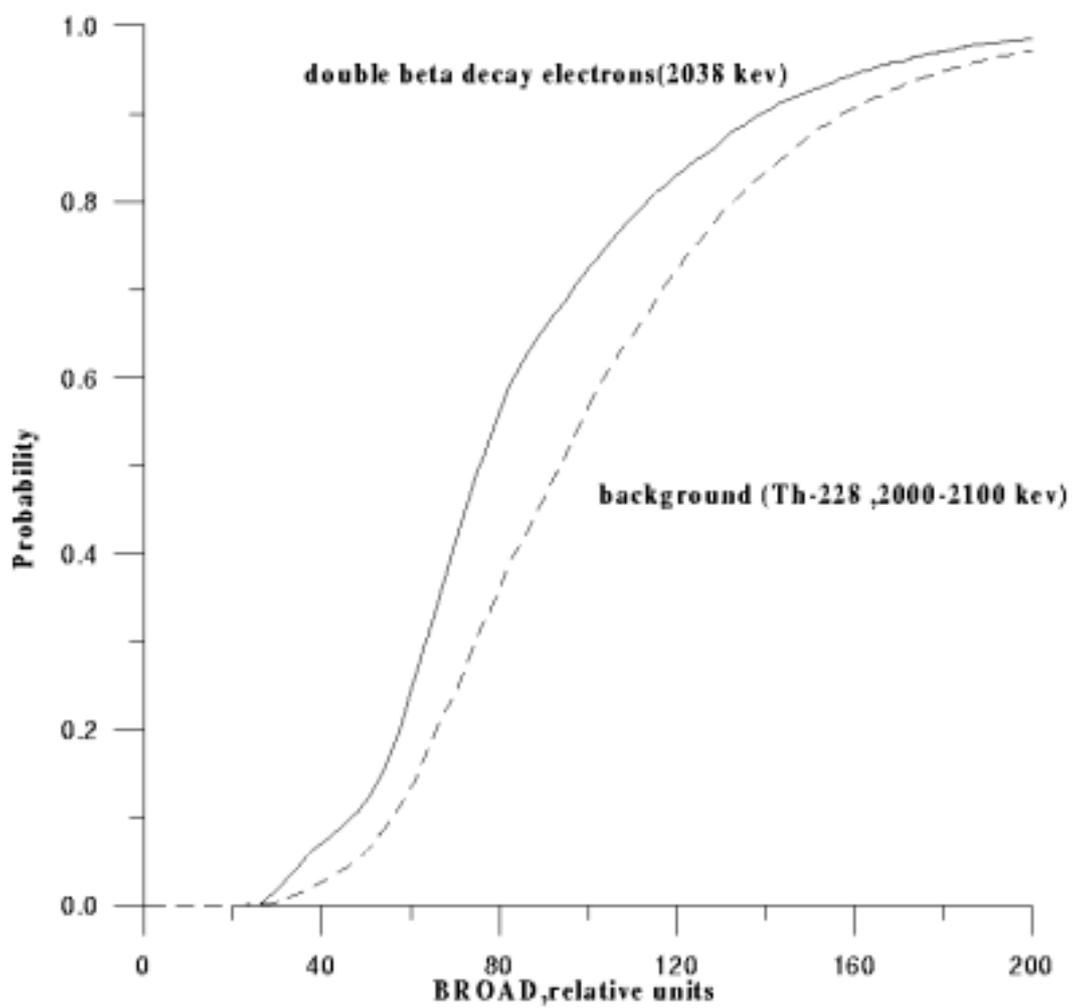
Fig.7. Integral density of probability for BROAD parameter

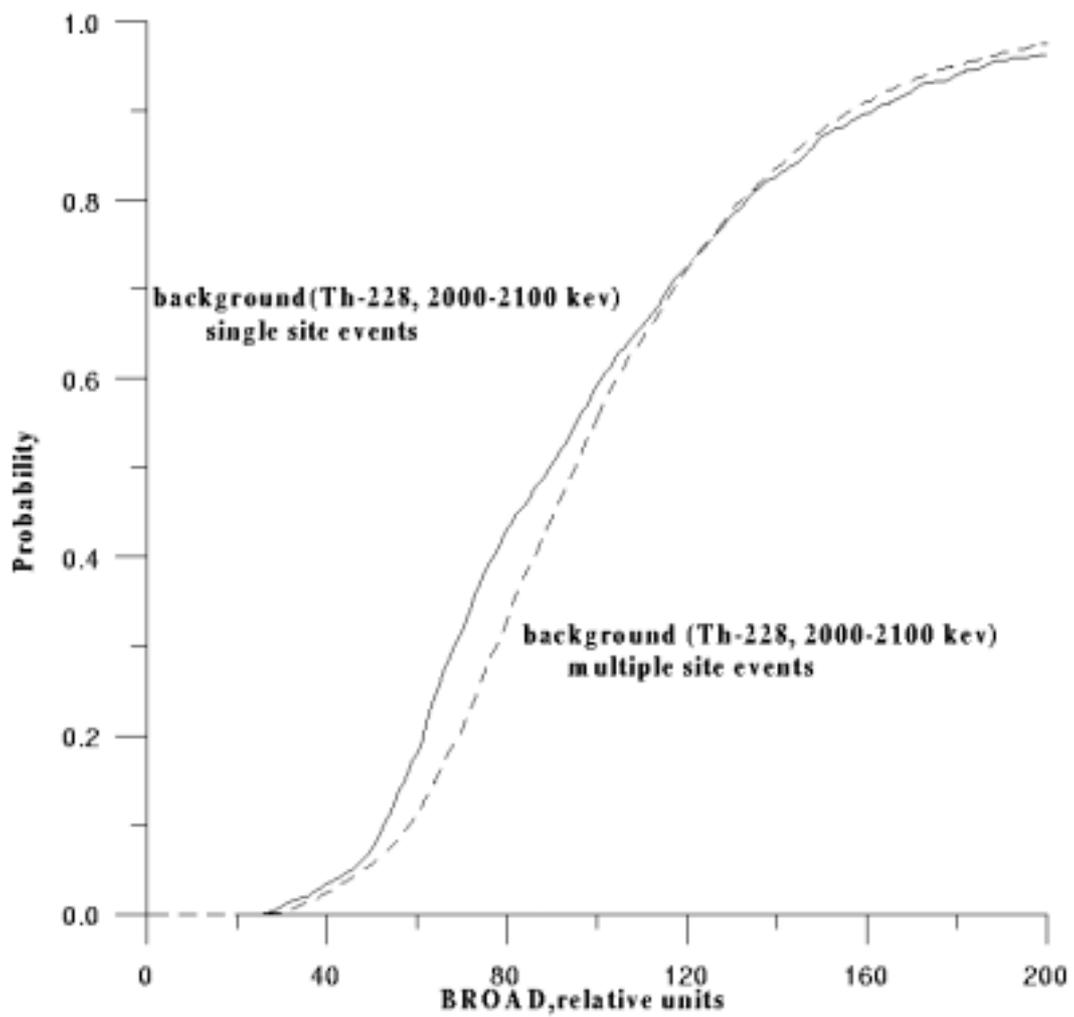

Fig.8. Integral density of probability for BROAD parameter

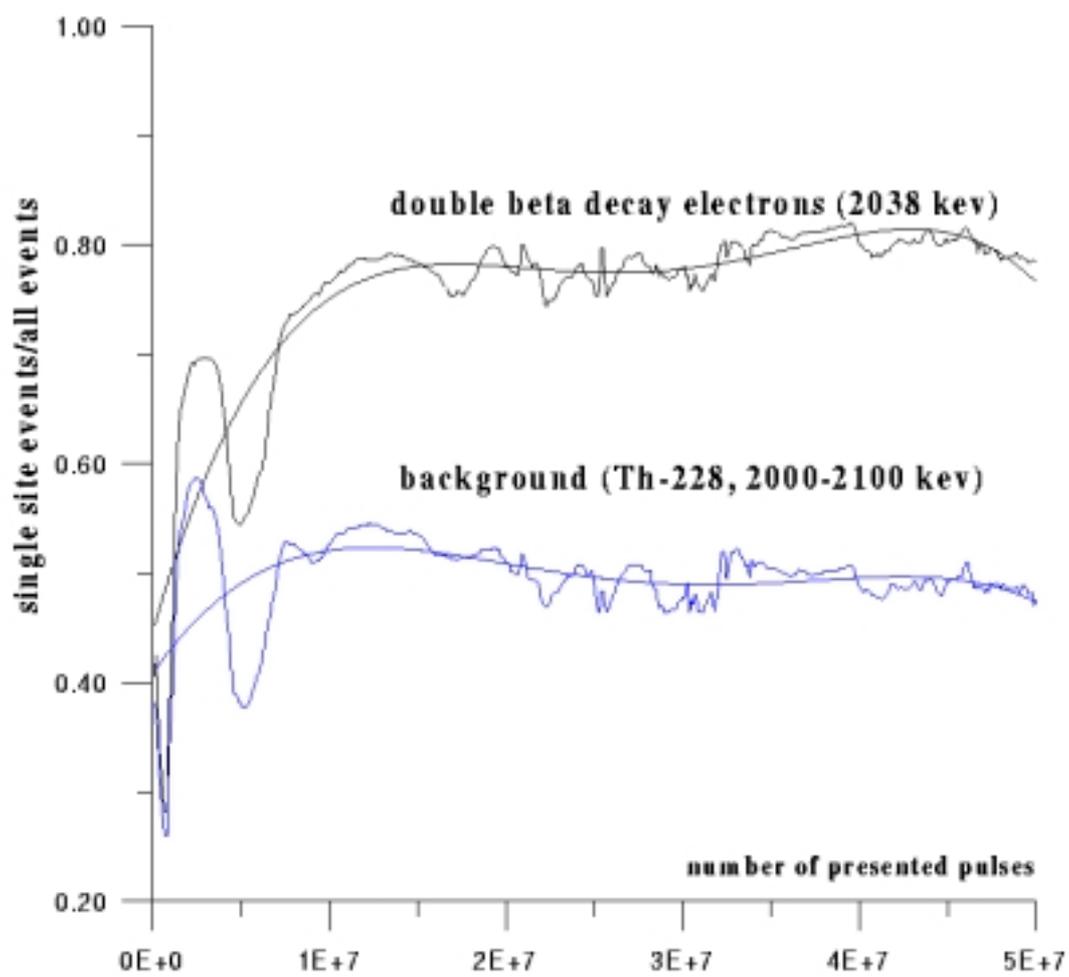

Fig.9 Dependence test parameter from training time

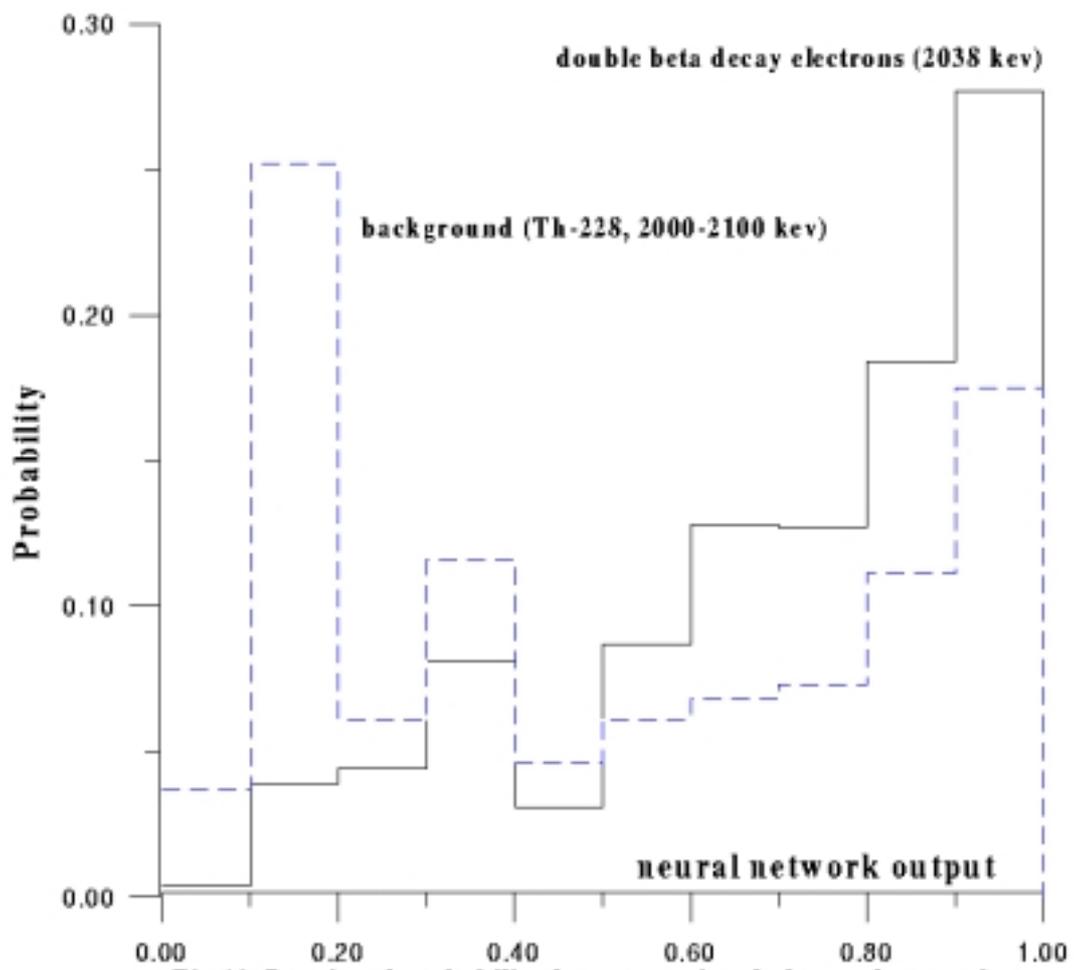

Fig.10. Density of probability for output signal of neural network

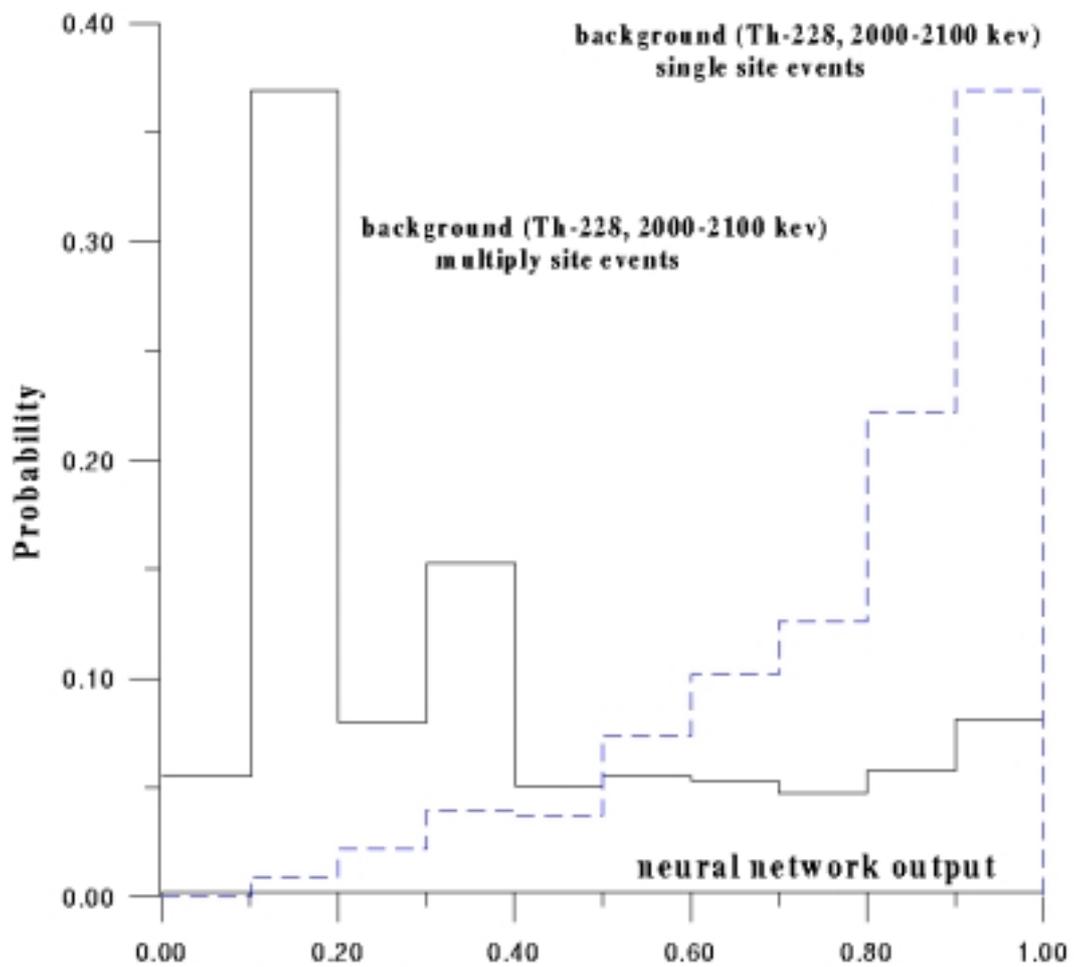

Fig.11. Density of probability for output signal of neural network

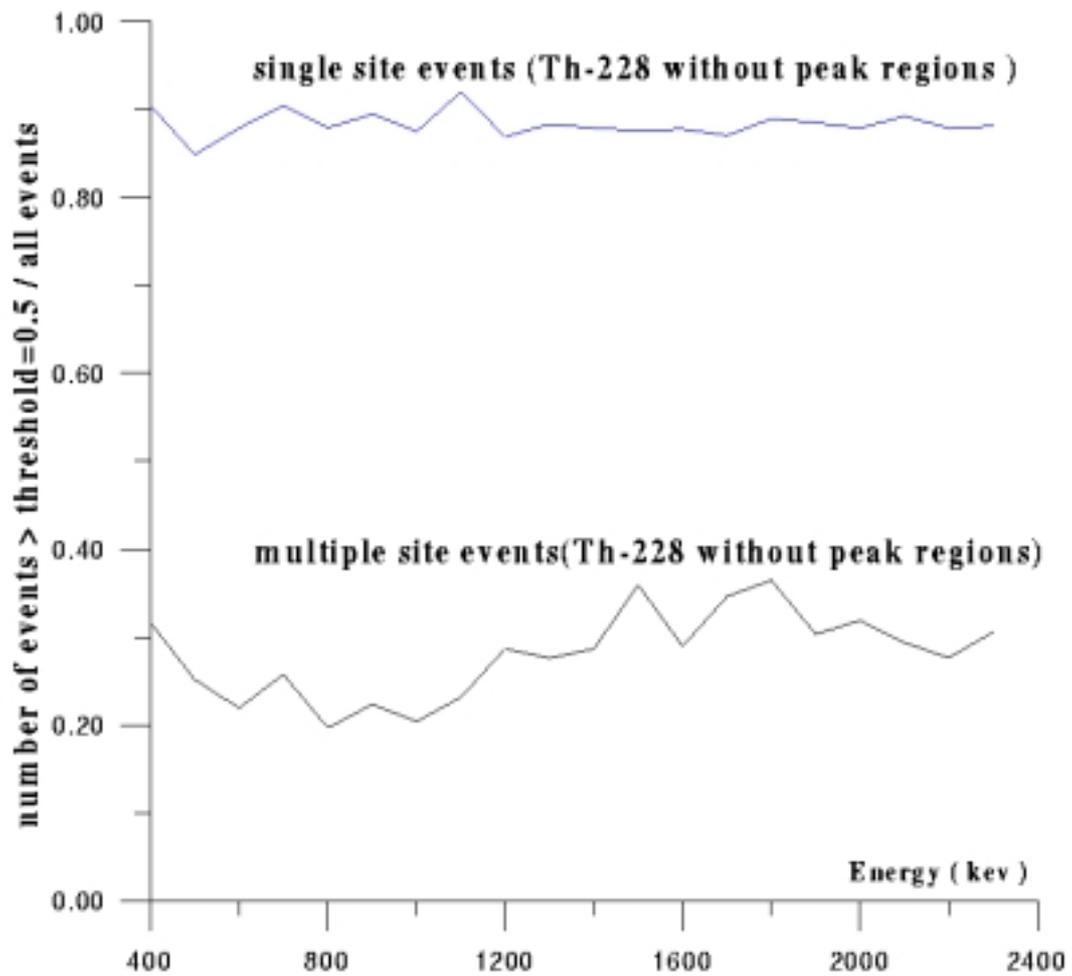

Fig.12 Function neural network output (separately for single site and multiple site events)

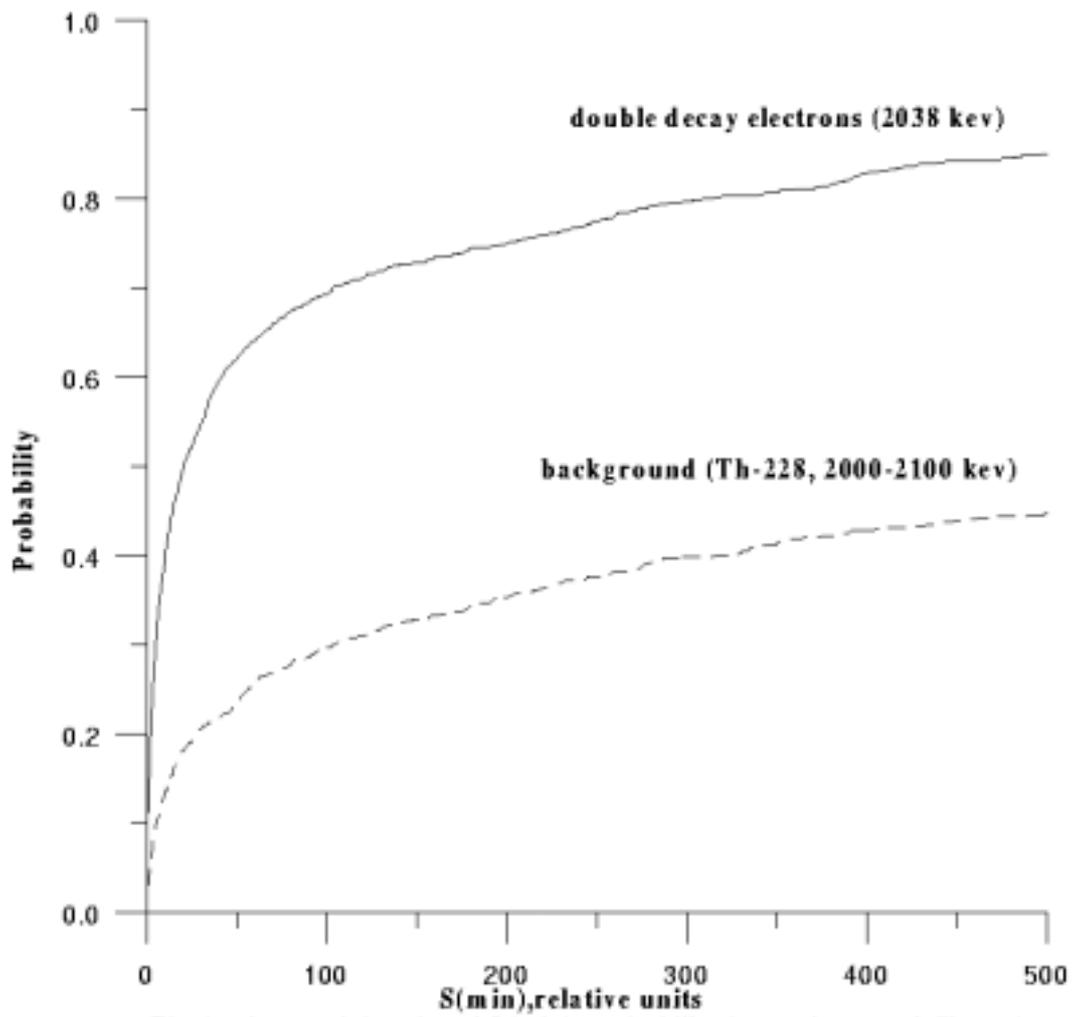

Fig.13. Integral density of S(min)-probability (neural network library)

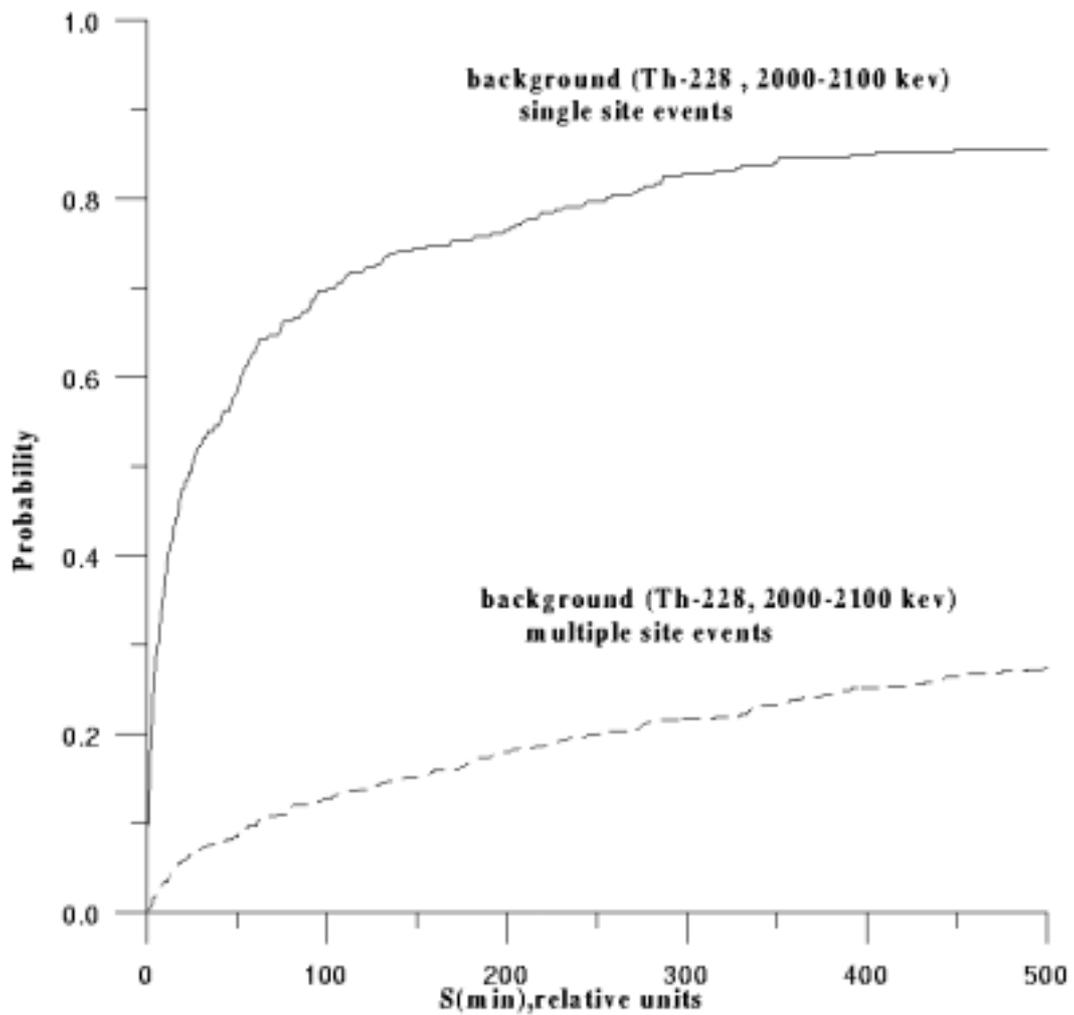

Fig.14. Integral density of S(min)-probability (neural network library)